\documentclass{amsart}
\usepackage{amssymb}
\usepackage{graphicx}
\usepackage{amscd}
\usepackage{amsmath}

\theoremstyle{plain}

\numberwithin{equation}{section}

\input{tcilatex}

\begin{document}
\title[Quantim Trajectories and Diffusion]{Quantum Trajectories, State
Diffusion and Time Asymmetric Eventum Mechanics }
\author{V P Belavkin}
\address{Mathematics Department, University of Nottingham, NG7 2RD, UK}
\email{vpb@maths.nott.ac.uk}
\urladdr{http://www.maths.nott.ac.uk/personal/vpb/}
\thanks{}
\date{}

\begin{abstract}
We show that the quantum stochastic Langevin model for continuous in time
measurements provides an exact formulation of the Heisenberg uncertainty
error-disturbance principle. Moreover, as it was shown in the 80's, this
Markov model induces all stochastic linear and non-linear equations of the
phenomenological "quantum trajectories" such as quantum state diffusion and
spontaneous localization by a simple quantum filtering method. Here we prove
that the quantum Langevin equation is equivalent to a Dirac type
boundary-value problem for the second-quantized input \textquotedblright
offer waves from future\textquotedblright\ in one extra dimension, and to a
reduction of the algebra of the consistent histories of past events to an
Abelian subalgebra for the \textquotedblleft trajectories of the output
particles\textquotedblright . This result supports the wave-particle duality
in the form of the thesis of Eventum Mechanics that everything in the future
is constituted by quantized waves, everything in the past by trajectories of
the recorded particles. We demonstrate how this time arrow can be derived
from the principle of quantum causality for nondemolition continuous in time
measurements.
\end{abstract}

\maketitle

\section{Introduction}

\begin{quote}
\textit{Quantum mechanics itself, whatever its interpretation, does not
account for the transition from `possible to the actual'}\textrm{\ }-\textrm{%
\ }Heisenberg.
\end{quote}

Schr\"{o}dinger believed that all problems of interpretation of quantum
mechanics including the above problem for time arrow should be formulated in
continuous time in the form of differential equations. He thought that the
quantum jump problem would have been resolved if quantum mechanics had been
made consistent with relativity theory of events and the time had been
treated appropriately as a future -- past boundary value problem of a
microscopic information dynamics. However Einstein and Heisenberg did not
believe this, each for his own reasons.

Although Schr\"{o}dinger did not succeed in finding the `true Schr\"{o}%
dinger equation' so he could formulate the boundary value problem for such
`eventum mechanics', the analysis of the phenomenological stochastic models
for quantum diffusions and spontaneous jumps proves that Schr\"{o}dinger was
right. We shall see that there exists indeed a boundary value problem for
the `true Schr\"{o}dinger equation' which corresponds to quantum jumps and
diffusive trajectories which is as continuous as Schr\"{o}dinger could have
wished, but it is not the usual Schr\"{o}dinger, but an ultrarelativistic
(massless) Dirac type boundary value problem in second quantization. However
Heisenberg was also right, as in order to take into account for these
transitions by filtering the actual past events simply as it is done in
classical statistics, the corresponding Dirac type boundary value problem
must be supplemented by future-past superselection rule for the total
algebra as it follows from the nondemolition causality principle \cite{Be94}%
. This principle demands the arrow of time, and it cannot be formulated in
the orthodox quantum mechanics as it involves infinitely many degrees of
freedom, and is yet unknown even in the quantum field theory.

Here we shall deal with quantum white noise models which allow us to
formulate the most general stochastic decoherence equation which was derived
in \cite{Be95a} from the unitary quantum Langevin equation. We shall start
with a simple quantum noise model and show that it allows us to prove the
`true Heisenberg principle' in the form of an uncertainty relation for
measurement errors and dynamical perturbations. The discovery of quantum
thermal noise and its white-noise approximations lead to a profound
revolution not only in modern physics but also in contemporary mathematics
comparable with the discovery of differential calculus by Newton (for a
feature exposition of this, accessible for physicists, see \cite{Gar91}, the
complete theory, which was mainly developed in the 80's \cite{Be80, HuPa84,
GaCo85, Be88a}, is sketched in the Appendix) Then we formulate the
corresponding boundary value problem of the Eventum Mechanics - the extended
quantum mechanics with a superselection causality rule in which there is a
place for microscopic events and trajectories. The dynamics of this
event-enhanced quantum mechanics is described by a one-parametric group of
unitary propagators on an extended Hilbert space, as in the conventional
quantum mechanics, however it is \emph{essentially irreversible}, as the
induced Heisenberg dynamics forms only a semigroup of \emph{invertible
endomorphisms} (but not of automorphisms!) in the positive arrow of time
chosen by the causality.

During the 90's many ``primary'' quantum theories appeared in the
theoretical and applied physics literature, in particular, the quantum state
diffusion theory \cite{GiPe92, GiPe93}, where a particular type of the
nonlinear quantum filtering stochastic equation have been used without even
a reference to the continuous measurements. The recent phenomenological
models for quantum trajectories in quantum optics \cite{Car93, WiMi93,
GoGr93, WiMi94, GoGr94, Car94} are also based on the stochastic solutions to
quantum jump equations, although the underlying boundary value problems of
eventum mechanics and the corresponding quantum stochastic filtering
equations of mathematical physics remain largely unknown in the general
physics. An exception occurred only in \cite{GoGr94, GGH95}, where our
quantum stochastic filtering theory which had been developed for these
purposes in the 80's, was well understood both at a macroscopic and
microscopic level. We complete this paper by formulating and discussing the
basic principles of the eventum mechanics as microscopic time asymmetric
information dynamics, which may include also classical mechanics, and which
is consistent with the quantum decoherence and quantum measurement.

\section{The true Heisenberg principle}

The first time-continuous model of the dynamical quantum measurement \cite%
{Be80} was motivated by analogy with the classical stochastic filtering
problem which obtains the prediction of the future for an unobservable
dynamical process $x\left( t\right) $ by time-continuous measuring of
another, observable process $y\left( t\right) $. Such problems were first
considered by Wiener and Kolmogorov who found the solutions in the form of
a\ causal spectral filter for a linear estimate $\hat{x}\left( t\right) $ of 
$x\left( t\right) $ which is optimal only in the stationary Gaussian case.
The complete solution of this problem was obtained by Stratonovich \cite%
{Str66} in 1958 who derived a stochastic filtering equation giving the
posterior expectations $\hat{x}\left( t\right) $\ of $x\left( t\right) $ in
the arbitrary Markovian pair $\left( x,y\right) $. This was really a break
through in the statistics of stochastic processes which soon found many
applications, in particular for solving the problems of stochastic control
under incomplete information (it is possible that this was one of the
reasons why the Russians were so successful in launching the rockets to the
Moon and other planets of the Solar system in 60s).

If $X\left( t\right) $ is an unobservable Heisenberg process, or vector of
such processes $X_{k}\left( t\right) $, $k=1,\ldots,d$ which might even have
no prior trajectories as the Heisenberg coordinate processes of a quantum
particle say, and $Y\left( t\right) $ is an actual observable quantum
processes, i.e. a sort of Bell's beable describing the vector trajectory $%
y\left( t\right) $ of the particle in a cloud chamber say, why don't we find
the posterior trajectories by deriving and solving a filtering equation for
the posterior expectations $\hat{x}\left( t\right) $ of $X\left( t\right) $
or any other \ function of $X\left( t\right) $, defining the posterior
trajectories $x\left( t,y_{0}^{t]}\right) $ in the same way as we do it in
the classical case? If we had a dynamical model in which such beables
existed as a nondemolition process, we could solve this problem simply by
conditioning as the statistical inference problem, predicting the future
knowing a history, i.e. a particular trajectory $y\left( r\right) $ up to
the time $t$. This problem was first considered and solved by finding a
nontrivial quantum stochastic model for the Markovian Gaussian pair $\left(
X,Y\right) $. It corresponds to a quantum open linear system with linear
output channel, in particular for a quantum oscillator matched to a quantum
transmission line \cite{Be80, Be85}. By studying this example, the
nondemolition condition 
\begin{equation*}
\left[ X_{k}\left( s\right) ,Y\left( r\right) \right] =0,\quad\text{ }\left[
Y\left( s\right) ,Y\left( r\right) \right] =0\quad\forall r\leq s
\end{equation*}
was first found, and this allowed the solution in the form of the causal
equation for $x\left( t,y_{0}^{t]}\right) =\left\langle X\left( t\right)
\right\rangle _{y_{0}^{t]}}$.

Let us describe this exact dynamical model of the causal nondemolition
measurement first in terms of quantum white noise for a quantum
nonrelativistic particle of mass $m$ which is conservative if not observed,
in a potential field $\phi$. But we shall assume that this particle is under
a time continuous indirect observation which is realized by measuring of its
Heisenberg position operators $Q^{k}\left( t\right) $ with additive random
errors $e^{k}\left( t\right) :$%
\begin{equation*}
Y^{k}\left( t\right) =Q^{k}\left( t\right) +e^{k}\left( t\right) ,\quad
k=1,\ldots,d.
\end{equation*}
We take the simplest statistical model for the error process $e\left(
t\right) $, the white noise model (the worst, completely chaotic error),
assuming that it is a classical Gaussian white noise given by the first
momenta 
\begin{equation*}
\left\langle e^{k}\left( t\right) \right\rangle =0,\quad\left\langle
e^{k}\left( s\right) e^{l}\left( r\right) \right\rangle
=\sigma_{e}^{2}\delta\left( s-r\right) \delta_{l}^{k}.
\end{equation*}
The components of measurement vector-process $Y\left( t\right) $ should be
commutative, satisfying the causal nondemolition condition with respect to
the noncommutative process $Q\left( t\right) $ (and any other Heisenberg
operator-process of the particle), this can be achieved by perturbing the
particle Newton-Ehrenfest equation: 
\begin{equation*}
m\frac{\mathrm{d}^{2}}{\mathrm{d}t^{2}}Q\left( t\right) +\nabla\phi\left(
Q\left( t\right) \right) =f\left( t\right) .
\end{equation*}
Here $f\left( t\right) $ is vector-process of Langevin forces $f_{k}$
perturbing the dynamics due to the measurement, which are also assumed to be
independent classical white noises 
\begin{equation*}
\left\langle f_{k}\left( t\right) \right\rangle =0,\quad\left\langle
f_{k}\left( s\right) f_{l}\left( r\right) \right\rangle
=\sigma_{f}^{2}\delta\left( s-r\right) \delta_{l}^{k}.
\end{equation*}
In classical measurement and filtering theory the white noises $e\left(
t\right) ,f\left( t\right) $ are usually considered independent, and the
intensities $\sigma_{e}^{2}$ and $\sigma_{f}^{2}$ can be arbitrary, even
zeros, corresponding to the ideal case of the direct unperturbing
observation of the particle trajectory $Q\left( t\right) $. However in
quantum theory corresponding to the standard commutation relations 
\begin{equation*}
Q\left( 0\right) =\mathrm{Q},\quad\frac{\mathrm{d}}{\mathrm{d}t}Q\left(
0\right) =\frac{1}{m}\mathrm{P},\quad\left[ \mathrm{Q}^{k},\mathrm{P}_{l}%
\right] =i\hbar\delta_{l}^{k}\mathrm{I}
\end{equation*}
the particle trajectories do not exist such that the measurement error $%
e\left( t\right) $ and perturbation force $f\left( t\right) $ should satisfy
a sort of uncertainty relation. This ``true Heisenberg principle'' had never
been mathematically proved before the discovery \cite{Be80} of quantum
causality in the form of nondemolition condition of commutativity of $%
Q\left( s\right) $, as well as any other process, the momentum $P\left(
t\right) =m\dot{Q}\left( t\right) $ say, with all $Y\left( r\right) $ for $%
r\leq s$. As we showed first in the linear case \cite{Be80, Be85}, and later
even in the most general case \cite{Be92b}, these conditions are fulfilled
if and only if $e\left( t\right) $ and $f\left( t\right) $ satisfy the
canonical commutation relations 
\begin{equation*}
\left[ e^{k}\left( r\right) ,e^{l}\left( s\right) \right] =0,\;\left[
e^{k}\left( r\right) ,f_{l}\left( s\right) \right] =\frac{\hbar}{i}%
\delta\left( r-s\right) \delta_{l}^{k},\;\left[ f_{k}\left( r\right)
,f_{l}\left( s\right) \right] =0.
\end{equation*}
From this it follows that the pair $\left( e,f\right) $ satisfy the
uncertainty relation $\sigma_{e}\sigma_{f}\geq\hbar/2$. This inequality
constitutes the precise formulation of the true Heisenberg principle for the
square roots $\sigma_{e}$ and $\sigma_{f}$ of the intensities of error $e$
and perturbation $f$: they are inversely proportional with the same
coefficient of proportionality, $\hbar/2$, as for the pair $\left( \mathrm{Q}%
,\mathrm{P}\right) $. Note that the canonical pair $\left( e,f\right) $
called quantum white noise cannot be considered classically, despite the
fact that each process $e$ and $f$ separately can. This is why we need a
quantum-field representation for the pair $\left( e,f\right) $, and the
corresponding quantum stochastic calculus. Thus, a generalized matrix
mechanics for the treatment of quantum open systems under continuous
nondemolition observation and the true Heisenberg principle was discovered
20 years ago only after the invention of quantum white noise in \cite{Be80}.
The nondemolition commutativity of $Y\left( t\right) $ with respect to the
Heisenberg operators of the open quantum system was later rediscovered for
the output of quantum stochastic fields in \cite{GaCo85}.

Let us outline the rigorous quantum stochastic model \cite{Be88, Be92b} for
a quantum particle of mass $m$ in a potential $\phi $ under indirect
observation of the positions $Q^{k}\left( t\right) $ by continual measuring
of continuous integral output processes $Y_{k}^{t}$. We will define the
output processes $Y_{k}^{t}$ as a quantum stochastic Heisenberg
transformation $Y_{k}^{t}=W\left( t\right) ^{\dagger }\left( \mathrm{I}%
\otimes \hat{y}_{k}^{t}\right) W\left( t\right) $ of the standard
independent Wiener processes $y_{k}^{t}=w_{k}^{t}$, $k=1,\ldots ,d$,
represented in Fock space $\mathcal{F}_{0}$ as the operators $\hat{y}%
_{k}^{t}=A_{-}^{k}\left( t\right) +A_{k}^{+}\left( t\right) \equiv \hat{w}%
_{k}^{t}$ on the Fock vacuum vector $\delta _{\varnothing }\in \mathcal{F}%
_{0}$ such that $w_{k}^{t}\simeq \hat{w}_{k}^{t}\delta _{\varnothing }$ (See
the notations and more about the quantum stochastic calculus in Fock space
in the Appendix 3). It has been shown in \cite{Be88, Be92b} that as such $%
W\left( t\right) $ one can take a time-continuous quantum stochastic unitary
evolution resolving an appropriate \emph{quantum stochastic Schr\"{o}dinger
equation }\ (See the equation (\ref{6.4}) below). It induces $Y_{k}^{t}$ as
the integrals of the following quantum stochastic output equations 
\begin{equation}
\text{\quad }\mathrm{d}Y_{k}^{t}=Q_{k}\left( t\right) \mathrm{d}t+\mathrm{d}%
\hat{w}_{k}^{t},  \label{6.1}
\end{equation}%
Here $Q_{k}\left( t\right) $ are the system Heisenberg operators $X\left(
t\right) =W\left( t\right) ^{\dagger }\left( \mathrm{X}\otimes I_{0}\right)
W\left( t\right) $ for $\mathrm{X}=\mathrm{Q}_{k}$ corresponding the system
position operators as $\mathrm{Q}^{k}=\left( 2\lambda _{k}\right) ^{-1}%
\mathrm{Q}_{k}$, where $\lambda _{k}$ are coupling constants defining in
general different accuracies of an indirect measurement in time of $\mathrm{Q%
}^{k}$, $I_{0}$ is the identity operator in the Fock space $\mathcal{F}_{0}$%
. This model coincides with the signal plus noise model given above with $%
e^{k}\left( t\right) $ being independent white noises of intensities
squaring $\sigma _{e}^{k}=\left( 2\lambda _{k}\right) ^{-1}$, represented as%
\begin{equation*}
e^{k}\left( t\right) =\frac{1}{2}\left( a_{k}^{+}+a_{-}^{k}\right) \left(
t\right) =\frac{1}{2\lambda _{k}}\frac{\mathrm{d}\hat{w}_{k}^{t}}{\mathrm{d}t%
},
\end{equation*}%
where $a_{k}^{+}\left( t\right) ,a_{-}^{k}\left( t\right) $ are the
canonical Bosonic creation and annihilation field operators, 
\begin{equation*}
\left[ a_{k}^{+}\left( s\right) ,a_{l}^{+}\left( t\right) \right] =0,\;\left[
a_{-}^{k}\left( s\right) ,a_{l}^{+}\left( t\right) \right] =\delta
_{l}^{k}\delta \left( t-s\right) ,\;\left[ a_{-}^{k}\left( s\right)
,a_{-}^{l}\left( t\right) \right] =0,
\end{equation*}%
defined as the generalized derivatives of the standard quantum Brownian
motions $A_{k}^{+}\left( t\right) $ and $A_{-}^{k}\left( t\right) $ in Fock
space $\mathcal{F}_{0}$. It was proved \ in \cite{Be88, Be92b} that $%
Y_{k}^{t}$ is a commutative nondemolition process with respect to the system
Heisenberg coordinate $Q\left( t\right) =W\left( t\right) ^{\dagger }\left( 
\mathrm{Q}\otimes I\right) W\left( t\right) $ and momentum $P\left( t\right)
=W\left( t\right) ^{\dagger }\left( \mathrm{P}\otimes I\right) W\left(
t\right) $ processes if they are perturbed by\ independent Langevin forces $%
f_{k}\left( t\right) $ of intensity $\sigma _{f}^{2}=\left( \lambda
_{k}\hbar \right) ^{2}$, the generalized derivatives of $f_{k}^{t}$, where $%
f_{k}^{t}=i\lambda _{k}\hbar \left( A_{-}^{k}-A_{k}^{+}\right) \left(
t\right) $: 
\begin{equation}
\mathrm{d}P_{k}\left( t\right) +\phi _{k}^{\prime }\left( Q\left( t\right)
\right) \mathrm{d}t=\mathrm{d}f_{k}^{t},\quad P_{k}\left( t\right) =m\frac{%
\mathrm{d}}{\mathrm{d}t}Q^{k}\left( t\right) .  \label{6.2}
\end{equation}%
Note that the quantum error operators $e_{k}^{t}=\hat{w}_{k}^{t}$ commute,
but they do not commute with the perturbing quantum force operators $%
f_{k}^{t}$ in Fock space due to the multiplication table 
\begin{equation*}
\left( \mathrm{d}e_{k}\right) ^{2}=I\mathrm{d}t,\quad \mathrm{d}f_{k}\mathrm{%
d}e_{l}=i\hbar \lambda _{k}I\delta _{k}^{l}\mathrm{d}t,\quad \mathrm{d}e_{k}%
\mathrm{d}f_{l}=-i\hbar \lambda _{l}I\delta _{l}^{k}\mathrm{d}t,\quad \left( 
\mathrm{d}f_{k}\right) ^{2}=\hbar ^{2}\lambda _{k}^{2}I\mathrm{d}t\text{.}
\end{equation*}%
This corresponds to the canonical commutation relations for their
generalized derivatives $e_{k}\left( t\right) $ of normalized intensity $%
\sigma _{e}^{2}=1$ and $f_{k}\left( t\right) $, so that the true Heisenberg
principle is fulfilled at the boundary $\sigma _{e}\sigma _{f}=\hbar \lambda
_{k}$. Thus our quantum stochastic model of nondemolition observation is the
minimal perturbation model for the given accuracies $\lambda _{k}$ of the
continual indirect measurement of the position operators $Q^{k}\left(
t\right) $ (the perturbations vanish when $\lambda _{k}=0$).

\subsubsection{Quantum state diffusion and filtering}

Let us introduce the quantum stochastic wave equation for the unitary
transformation $\Psi _{0}\left( t\right) =W\left( t\right) \Psi _{0}$
inducing Heisenberg dynamics which is described by the quantum Langevin
equation (\ref{6.2}) with white noise perturbation in terms. This equation
is well understood in terms of the generalized derivatives 
\begin{equation*}
f_{k}\left( t\right) =\lambda _{k}\frac{\hbar }{i}\left(
a_{k}^{+}-a_{-}^{k}\right) \left( t\right) =\frac{\mathrm{d}f_{k}^{t}}{%
\mathrm{d}t}
\end{equation*}%
of the standard quantum Brownian motions $A_{k}^{+}\left( t\right) $ and $%
A_{-}^{k}\left( t\right) $ defined by the commutation relations 
\begin{equation*}
\left[ A_{k}^{+}\left( s\right) ,A_{l}^{+}\left( t\right) \right] =0,\;\left[
A_{-}^{k}\left( s\right) ,A_{l}^{+}\left( t\right) \right] =\left( t\wedge
s\right) \delta _{l}^{k},\;\left[ A_{-}^{k}\left( s\right) ,A_{-}^{l}\left(
t\right) \right] =0
\end{equation*}%
in Fock space $\mathcal{F}_{0}$ ($t\wedge s=\min \left\{ s,t\right\} $). \
The corresponding quantum stochastic differential equation 
\begin{equation}
\mathrm{d}\Psi _{0}\left( t\right) +\left( \mathrm{K}\otimes I\right) \Psi
_{0}\left( t\right) \mathrm{d}t=\frac{i}{\hbar }\left( \mathrm{Q}^{k}\otimes 
\mathrm{d}f_{k}^{t}\right) \Psi _{0}\left( t\right) ,\;\Psi _{0}\left(
0\right) =\Psi _{0}  \label{6.4}
\end{equation}%
for the probability amplitude in $\mathfrak{h}\otimes \mathcal{F}_{0}$ is a
particular case 
\begin{equation*}
\mathrm{L}_{k}^{-\dagger }=-\mathrm{L}_{+}^{k},\quad \mathrm{L}%
_{+}^{k}=\left( \lambda \mathrm{Q}\right) ^{k}\equiv \mathrm{L}^{k}
\end{equation*}%
of the general quantum diffusion wave equation%
\begin{equation*}
\mathrm{d}\Psi _{0}\left( t\right) +\left( \mathrm{K}\otimes I\right) \Psi
_{0}\left( t\right) \mathrm{d}t=\left( \mathrm{L}_{+}^{k}\otimes \mathrm{d}%
A_{k}^{+}+\mathrm{L}_{k}^{-}\otimes \mathrm{d}A_{-}^{k}\right) \left(
t\right) \Psi _{0}\left( t\right)
\end{equation*}%
which describes the unitary evolution in $\mathfrak{h}\otimes \mathcal{F}%
_{0} $ if $\mathrm{K}=\frac{i}{\hbar }\mathrm{H}-\frac{1}{2}\mathrm{L}%
_{k}^{-}\mathrm{L}_{+}^{k}$, where $\mathrm{H}^{\dagger }=\mathrm{H}$ is the
evolution Hamiltonian for the system in $\mathfrak{h}$. Using the quantum It%
\^{o} formula, see the Appendix 3, it was proven in \cite{Be88, Be92b} that
it is equivalent to the Langevin equation 
\begin{eqnarray}
\mathrm{d}Z\left( t\right) &=&\left( g\left( ZL+L^{\dagger }Z\right)
+L^{\dagger }ZL-K^{\dagger }Z-ZK\right) \left( t\right) \mathrm{d}t  \notag
\\
&&+\left( gZ+L^{\dagger }Z-ZL\right) \left( t\right) \mathrm{d}A_{-}+\left(
gZ+ZL-L^{\dagger }Z\right) \left( t\right) \mathrm{d}A^{+}  \label{6.5}
\end{eqnarray}%
for any quantum stochastic Heisenberg process 
\begin{equation*}
Z\left( t,\mathrm{X},g\right) =W\left( t\right) ^{\dagger }\left( \mathrm{X}%
\otimes \exp \left[ \int_{0}^{t}\left( g^{k}\left( r\right) \mathrm{d}\hat{w}%
_{k}^{r}-\frac{1}{2}g\left( r\right) ^{2}\mathrm{d}r\right) \right] \right)
W\left( t\right) ,
\end{equation*}%
where $g^{k}\left( t\right) $ are a test function for the output process $%
w_{k}^{t}$ and 
\begin{equation*}
K\left( t\right) =W\left( t\right) ^{\dagger }\left( \mathrm{K}\otimes 
\mathrm{I}\right) W\left( t\right) ,\quad L^{k}\left( t\right) =W\left(
t\right) ^{\dagger }\left( \mathrm{L}^{k}\otimes \mathrm{I}\right) W\left(
t\right) .
\end{equation*}%
The Langevin equation (\ref{6.2}) for the system coordinate $Q\left(
t\right) =W\left( t\right) ^{\dagger }\left( \mathrm{Q}\otimes I\right)
W\left( t\right) $ corresponding to $\mathrm{X}=\mathrm{Q}$, $g=0$ with the
generating exponent $Y\left( t,g\right) =Z\left( t,\mathrm{I},g\right) $ for
the output processes $Y_{k}^{t}=W\left( t\right) ^{\dagger }\left( \mathrm{I}%
\otimes \hat{w}_{k}^{t}\right) W\left( t\right) $ corresponding to $\mathrm{X%
}=\mathrm{I}$ follow straightforward in the case $\mathrm{L}=\lambda \mathrm{%
Q}$, $\mathrm{H}=\mathrm{P}^{2}/2m+\phi \left( \mathrm{Q}\right) $.

In the next section we shall show that this unitary evolution is the
interaction picture for a unitary group evolution $U^{t}$ corresponding to a
Dirac type boundary value problem for a generalized Schr\"{o}dinger equation
in an extended product Hilbert space $\mathfrak{h}\otimes \mathcal{G}$. Here
we prove that the quantum stochastic evolution (\ref{6.4}) in $\mathfrak{h}%
\otimes \mathcal{F}_{0}$ coincides with the quantum state diffusion in $%
\mathfrak{h}$ if it is considered only for the initial product states $\psi
\otimes \delta _{\varnothing }$ with $\delta _{\varnothing }$ being the Fock
vacuum state vector in $\mathcal{F}_{0}$.

\emph{Quantum state diffusion} is a nonlinear, nonunitary, irreversible
stochastic form of quantum mechanics with classical histories, called
sometimes also quantum mechanics with trajectories. It was put forward by
Gisin and Percival \cite{GiPe92, GiPe93} in the early 90's as a new, \emph{%
primary} quantum theory which includes the diffusive reduction process into
the wave equation for pure quantum states. It has been criticized, quite
rightly, as an incomplete theory which does not satisfy the linear
superposition principle for the waves, and for not explaining the origin of
irreversible dissipativity which is build into the equation `by hands'. In
fact the `primary' equation had been derived even earlier as the posterior
state diffusion equation for pure states $\psi _{\omega }=\psi \left( \omega
\right) /\left\Vert \psi \left( \omega \right) \right\Vert $ from the linear
unitary quantum diffusion equation (\ref{6.4}) by the following method as a
particular type of the general quantum filtering equation in \cite{Be88,
Be89b, BeSt92}. Here we shall show only how to derive the corresponding
stochastic linear decoherence equation \{\ref{5.1}) for $\psi \left(
t,w\right) =V\left( t,w\right) \psi $ when all the independent increment
processes $y_{k}^{t}$ are of the diffusive type $y_{k}^{t}=w_{k}^{t}$: 
\begin{equation}
\mathrm{d}\psi \left( t,\omega \right) +\left( \frac{i}{\hbar }\mathrm{H}+%
\frac{1}{2}\mathrm{Q}^{k}\lambda _{k}^{2}\mathrm{Q}^{k}\right) \psi \left(
t,\omega \right) \mathrm{d}t=\left( \lambda \mathrm{Q}\right) ^{k}\psi
\left( t,\omega \right) \mathrm{d}w_{k}^{t}.  \label{6.6}
\end{equation}%
Note that the resolving stochastic propagator $V\left( t,\omega \right) $
for this equation defines the isometries 
\begin{equation*}
V\left( t\right) ^{\dagger }V\left( t\right) =\int V\left( t,\omega \right)
^{\dagger }V\left( t,\omega \right) \mathrm{d}\mu =1
\end{equation*}%
of the system Hilbert space $\mathfrak{h}$ into the Wiener Hilbert space $%
L_{\mu }^{2}$ of square integrable functionals of the diffusive trajectories 
$\omega =\left\{ w\left( t\right) \right\} $ with respect to the standard
Gaussian measure $\mu =\mathsf{P}_{w}$ if $\mathrm{K}+\mathrm{K}^{\dagger }=%
\mathrm{L}^{\dagger }\mathrm{L}$.

Let us represent these Wiener processes in the equation by operators $%
e_{k}^{t}=A_{k}^{+}+A_{-}^{k}$ on the Fock space vacuum $\delta
_{\varnothing }$ using the unitary equivalence $w_{k}^{t}\simeq
e_{k}^{t}\delta _{\varnothing }$ in the notation explained in the Appendix
3. Then the corresponding quantum stochastic equation 
\begin{equation*}
\mathrm{d}\hat{\psi}\left( t\right) +\mathrm{K}\hat{\psi}\left( t\right) 
\mathrm{d}t=\left( \lambda \mathrm{Q}\right) ^{k}\mathrm{d}e_{k}^{t}\hat{\psi%
}\left( t\right) ,\quad \hat{\psi}\left( 0\right) =\psi \otimes \delta
_{\varnothing },\psi \in \mathfrak{h}
\end{equation*}%
coincides with the quantum diffusion Schr\"{o}dinger equation (\ref{6.4}) on
the same initial product-states $\Psi _{0}=\psi \otimes \delta _{\varnothing
}$. Indeed, as it was noted in \cite{Be92b}, due to the adaptedness 
\begin{equation*}
\hat{\psi}\left( t\right) =\hat{\psi}^{t}\otimes \delta _{\varnothing
},\quad \Psi _{0}\left( t\right) =\Psi _{0}^{t}\otimes \delta _{\varnothing }
\end{equation*}%
both right had sides of these equations coincide on future vacuum $\delta
_{\varnothing }$ if $\hat{\psi}^{t}=\Psi _{0}^{t}$ as 
\begin{eqnarray*}
\left( \lambda \mathrm{Q}\right) ^{k}\mathrm{d}e_{k}^{t}\hat{\psi}\left(
t\right) &=&\left( \mathrm{L}^{k}\mathrm{d}A_{k}^{+}+\mathrm{L}_{k}^{\dagger
}\mathrm{d}A_{-}^{k}\right) \left( \hat{\psi}^{t}\otimes \delta
_{\varnothing }\right) =\mathrm{L}^{k}\hat{\psi}^{t}\otimes \mathrm{d}%
A_{k}^{+}\delta _{\varnothing } \\
\frac{i}{\hbar }\mathrm{Q}^{k}\mathrm{d}f_{k}^{t}\Psi _{0}\left( t\right)
&=&\left( \mathrm{L}^{k}\mathrm{d}A_{k}^{+}-\mathrm{L}_{k}\mathrm{d}%
A_{-}^{k}\right) \left( \Psi _{0}^{t}\otimes \delta _{\varnothing }\right) =%
\mathrm{L}^{k}\Psi _{0}^{t}\otimes \mathrm{d}A_{k}^{+}\delta _{\varnothing }
\end{eqnarray*}%
(the annihilation processes $A_{-}^{k}$ are zero on the vacuum $\delta
_{\varnothing }$). By virtue of the coincidence of the initial data $\hat{%
\psi}^{0}=\psi =\Psi _{0}^{0}$ this proves that $\hat{\psi}\left( t\right) $
is also the solution $\Psi _{0}\left( t\right) =W\left( t\right) \left( \psi
\otimes \delta _{\varnothing }\right) $ of (\ref{6.4}) for all $t>0$: 
\begin{equation*}
\mathrm{d}\hat{\psi}\left( t\right) +\mathrm{K}\hat{\psi}\left( t\right) 
\mathrm{d}t=\frac{i}{\hbar }\left( \mathrm{Q}^{k}\otimes \mathrm{d}%
f_{k}^{t}\right) \hat{\psi}\left( t\right) ,\quad \hat{\psi}\left( 0\right)
=\psi \otimes \delta _{\varnothing },\psi \in \mathfrak{h}.
\end{equation*}
Note that the latter equation can be written as a classical stochastic Schr%
\"{o}dinger equation 
\begin{equation}
\mathrm{d}\tilde{\psi}\left( t,\upsilon \right) +\left( \frac{i}{\hbar }%
\mathrm{H}+\frac{1}{2}\mathrm{Q}^{k}\lambda _{k}^{2}\mathrm{Q}^{k}\right) 
\tilde{\psi}\left( t,\upsilon \right) \mathrm{d}t=\frac{i}{\hbar }\left( 
\mathrm{Q}^{k}\otimes \mathrm{d}u_{k}^{t}\right) \tilde{\psi}\left(
t,\upsilon \right)  \label{6.3}
\end{equation}
for the Fourier-Wiener transform $\tilde{\psi}\left( \upsilon \right) $ of $%
\psi \left( \omega \right) $ in terms of another Wiener processes $u_{k}^{t}$
having Fock space representations $f_{k}^{t}\delta _{\varnothing }$. The
stochastic unitary propagator $\check{W}\left( t,\upsilon \right) $
resolving this equation on another probability space of $\upsilon =\left\{
u\left( t\right) \right\} $ is given by the corresponding transformation $%
\check{W}=\tilde{V}$ of the nonunitary stochastic propagator $V\left(
t,\omega \right) $ for the equation (\ref{6.6}).

Thus the stochastic decoherence equation (\ref{6.6}) for the continuous
observation of the position of a quantum particle with $\mathrm{H}=\frac{1}{%
2m}\mathrm{P}^{2}+\phi \left( \mathrm{Q}\right) $ was derived for the
unitary quantum stochastic evolution as an example of the general
decoherence equation (\ref{5.1}) which was obtained in this way in \cite%
{Be88}. It was explicitly solved in \cite{Be88, Be89b, BeSt92} for the case
of linear and quadratic potentials $\phi $, and it was shown that this
solution coincides with the optimal quantum linear filtering solution
obtained earlier in \cite{Be80, Be85} if the initial wave packet is Gaussian.

The nonlinear stochastic posterior equation for this particular case was
derived independently by Diosi \cite{Dio88} and (as an example) in \cite%
{Be88, Be89b}. It has the following form 
\begin{equation*}
\mathrm{d}\psi _{w}\left( t\right) +\left( \frac{i}{\hbar }\mathrm{H}+\frac{1%
}{2}\widetilde{\mathrm{Q}}^{k}\left( t\right) \lambda _{k}^{2}\widetilde{%
\mathrm{Q}}^{k}\left( t\right) \right) \psi _{w}\left( t\right) \mathrm{d}%
t=\lambda _{k}\widetilde{\mathrm{Q}}^{k}\left( t\right) \psi _{w}\left(
t\right) \mathrm{d}\tilde{w}_{k}^{t},
\end{equation*}%
where $\widetilde{\mathrm{Q}}\left( t\right) =\mathrm{Q}-\hat{q}\left(
t\right) $ with $\hat{q}^{k}\left( t\right) $ defined as the multiplication
operators by the components $q^{k}\left( t,w\right) =\psi _{w}^{\dagger
}\left( t\right) \mathrm{Q}^{k}\left( t\right) \psi _{w}\left( t\right) $ of
the posterior expectation (statistical prediction) of the coordinate $%
\mathrm{Q}$, and 
\begin{equation*}
\mathrm{d}\tilde{w}_{k}^{t}=\mathrm{d}w_{k}^{t}-2\left( \lambda \hat{q}%
\right) ^{k}\left( t\right) \mathrm{d}t=\mathrm{d}y_{k}^{t}-\hat{x}%
_{k}\left( t\right) \mathrm{d}t,\quad \hat{x}_{k}\left( t\right) =2\left(
\lambda \hat{q}\right) ^{k}\left( t\right) .
\end{equation*}%
Note that the innovating output processes $\tilde{w}_{k}^{t}$ are also
standard Wiener processes with respect to the output probability measure $%
\mathrm{d}\tilde{\mu}=\lim \Pr \left( t,\mathrm{d}\omega \right) $, but not
with respect to the Wiener probability measure $\mu =\Pr \left( 0,\mathrm{d}%
\omega \right) $ for the input noise $w_{k}^{t}$.

Let us give the explicit solution of this stochastic wave equation for the
free particle ($\phi=0$) in one dimension and the stationary Gaussian
initial wave packet which was found in \cite{Be88, Be89b, BeSt92}. One can
show \cite{ChSt92, Kol95} that the nondemolition observation of such
particle is described by filtering of quantum noise which results in the
continual collapse of any wave packet to the Gaussian stationary one
centered at the posterior expectation $q\left( t,w\right) $ with finite
dispersion $\left\| \left( \hat{q}\left( t\right) -\mathrm{Q}\right)
\psi_{\omega}\left( t\right) \right\| ^{2}\rightarrow2\lambda\left(
\hbar/m\right) ^{1/2}$. This center can be found from the linear Newton
equation 
\begin{equation*}
\frac{\mathrm{d}^{2}}{\mathrm{d}t^{2}}z\left( t\right) +2\kappa \frac{%
\mathrm{d}}{\mathrm{d}t}z\left( t\right) +2\kappa^{2}z\left( t\right)
=-g\left( t\right) ,
\end{equation*}
for the deviation process $z\left( t\right) =q\left( t\right) -x\left(
t\right) $, where $x\left( t\right) $ is an expected trajectory of the
output process (\ref{6.1}) with $z\left( 0\right) =q_{0}-x\left( 0\right) $, 
$z^{\prime}\left( 0\right) =v_{0}-x^{\prime}\left( 0\right) $. Here $%
\kappa=\lambda\left( \hbar/m\right) ^{1/2}$ is the decay rate which is also
the frequency of effective oscillations, $q_{0}=\left\langle \hat {x}%
\right\rangle $, $v_{0}=\left\langle \hat{p}/m\right\rangle $ are the
initial expectations and $g\left( t\right) =x^{\prime\prime}\left( t\right) $
is the effective gravitation for the particle in the moving framework of $%
x\left( t\right) $. The solution to the above equation illustrate the
continuous collapse $z\left( t\right) \rightarrow0$ of the posterior
trajectory $q\left( t\right) $ towards a linear trajectory $x\left( t\right) 
$. The posterior position expectation $q\left( t\right) $ in the absence of
effective gravitation, $x^{\prime\prime}\left( t\right) =0$, for the linear
trajectory $x\left( t\right) =ut-q$ collapses to the expected input
trajectory $x\left( t\right) $ with the rate $\kappa =\lambda\left(
\hbar/m\right) ^{1/2}$, remaining not collapsed, $q_{0}\left( t\right)
=v_{0}t$ in the framework where $q_{0}=0$, only in the classical limit $%
\hbar/m\rightarrow0$ or absence of observation $\lambda=0$. This is the
graph of 
\begin{equation*}
q_{0}\left( t\right) =v_{0}t,\quad q\left( t\right) =ut+e^{-\kappa t}\left(
q\cos\kappa t+\left( q+\kappa^{-1}\left( v_{0}-u\right) \right) \sin\kappa
t\right) -q
\end{equation*}
obtained as $q\left( t\right) =x\left( t\right) +z\left( t\right) $ by
explicit solving of the second order linear equation for $z\left( t\right) $.

\section{The eventum mechanics realization}

Finally, let us describe the \emph{Eventum Mechanics} underlying all quantum
diffusion and more general quantum noise Langevin models of information
dynamics. We shall see that all such phenomenological models exactly
correspond to Dirac type boundary value problems for a Poisson flow of
independent quantum particles interacting with the quantum system under the
observation at the boundary $r=0$ of the half line $\mathbb{R}_{+}$ in an
additional dimension. The second-quantized massless Dirac equation (\ref{6.7}%
) with corresponding boundary condition (\ref{6.8}), together with the
quantum causality (or nondemolition) superselection rule, is the essence of
the Eventum Mechanics, the new, extended quantum mechanics in which there is
a place for the phenomenological events such as quantum trajectories and
spontaneous localizations. One can think of the coordinate $r>0$ being
perpendicular to the quantum target membrane of a scattering measuring
device, or an extra dimension coordinate as a physical realization of
localizable time in our m-brane universe. At least it is so for any free
evolution Hamiltonian $\varepsilon \left( p\right) >0$ of the incoming
quantum particles in the ultrarelativistic limit $\left\langle
p\right\rangle \rightarrow -\infty $ such that the average velocity in an
initial state is a finite constant, $c=\left\langle \varepsilon ^{\prime
}\left( p\right) \right\rangle \rightarrow 1$ say, see in details on this
limit of an idealized rigid boundary measurement schemes in \cite{Be00,
Be01a, BeKo01}. Thus we are going to solve the following problem of the
derivation of time continuous information dynamics as the microscopic
foundation problem for quantum trajectories, individual decoherence, state
diffusion, or permanent reduction theories:

\textbf{The dynamical quantum measurement problem.}

\emph{Let }$V\left( t,\omega \right) =V\left( t,w_{0}^{t]}\right) ,t\in 
\mathbb{R}_{+}$\emph{\ be a reduction family} \emph{of isometries on} $%
\mathfrak{h}$\ \emph{into} $\mathfrak{h}\otimes L_{\mu }^{2}$\emph{\
resolving the state diffusion equation (\ref{6.6}) with respect to the input
probability measure }$\mu =\mathsf{P}_{w}$\emph{\ for the standard Wiener
noises }$w_{k}^{t}$\emph{, defining the classical means} 
\begin{equation*}
\mathsf{M}\left[ gV\left( t\right) ^{\dagger }\mathrm{B}V\left( t\right) %
\right] =\int g\left( w_{0}^{t]}\right) V\left( t,w_{0}^{t]}\right)
^{\dagger }\mathrm{B}V\left( t,w_{0}^{t]}\right) \mathrm{d}\mathsf{P}_{w}.
\end{equation*}%
\emph{\ Find a triple }$\left( \mathcal{G},\mathfrak{A},\Phi \right) $ \emph{%
consisting of a Hilbert space} $\mathcal{G=G}_{-}\otimes \mathcal{G}_{+}$ 
\emph{embedding the Wiener Hilbert space }$L_{\mu }^{2}$\emph{\ by an
isometry into }$\mathcal{G}_{+}$\emph{, an algebra }$\mathfrak{A=A}%
_{-}\otimes \mathfrak{A}_{+}$ \emph{on $\mathcal{G}$ with an Abelian
subalgebra }$\mathfrak{A}_{-}$\emph{\ generated by a compatible continuous
family }$Y_{-\infty }^{0]}=\left\{ Y_{k}^{s},k=1,\ldots ,d,s\leq 0\right\} $%
\emph{\ } \emph{of observables on }$\mathcal{G}_{-}$\emph{, and a
state-vector }$\Phi ^{\circ }=\Phi _{-}^{\circ }\otimes \Phi _{+}^{\circ
}\in \mathcal{G}$\emph{\ such that there exists a time continuous unitary
group }$U^{t}$\emph{\ on }$\mathcal{H=}\mathfrak{h}\otimes \mathcal{G}$\emph{%
,\ inducing a semigroup of endomorphisms }$\mathfrak{B}\ni B\mapsto
U^{-t}BU^{t}\in \mathfrak{B}$\emph{, which represents this reduction on the
product algebra $\mathfrak{B=}\mathcal{B}\left( \mathfrak{h}\right) \otimes 
\mathfrak{A}$ as} 
\begin{equation*}
\mathsf{M}\left[ gV\left( t\right) ^{\dagger }\mathrm{B}V\left( t\right) %
\right] =\pi ^{t}\left( \hat{g}_{-t}\otimes \mathrm{B}\right) .
\end{equation*}%
\emph{Here} $\pi ^{t}$\emph{\ is the quantum conditional expectation} 
\begin{equation*}
\pi ^{t}\left( \hat{g}_{-t}\otimes \mathrm{B}\right) =\left( \mathrm{I}%
\otimes \Phi ^{\circ }\right) ^{\dagger }U^{-t}\left( \mathrm{B}\otimes
g_{-t}\left( Y_{-t}^{0]}\right) \right) U^{t}\left( \mathrm{I}\otimes \Phi
^{\circ }\right) ,
\end{equation*}%
\ \emph{which provides the dynamical realization of the reduction as the
statistically causal inference about any }$\mathrm{B}\in \mathcal{B}\left( 
\mathfrak{h}\right) $ \emph{with respect to the algebra }$\mathfrak{A}_{-}$%
\emph{\ of the functionals }$\hat{g}_{-t}=g_{-t}\left( Y_{-t}^{0]}\right) $%
\emph{\ of }$Y_{-t}^{-0]}=\left\{ Y_{\cdot }^{s}:s\in (-t,0]\right\} $\emph{%
, all commuting on $\mathcal{G}$, representing the shifted measurable
functionals }$g_{-t}\left( y_{-t}^{0]}\right) =g\left( y_{0}^{t]}\right) $%
\emph{\ of }$y_{0}^{t]}=\left\{ y_{\cdot }^{r}:r\in (0,t]\right\} $\emph{\
for each }$t>0$\emph{\ in the center }$\mathfrak{C}$ \emph{\ of the algebra }%
$\mathfrak{B}$\emph{.}

\textbf{Solution:} We have already dilated the state diffusion equation (\ref%
{6.6}) to a quantum stochastic unitary evolution $W\left( t\right) $
resolving the quantum stochastic Schr\"{o}dinger equation (\ref{6.4}) on the
system Hilbert $\mathfrak{h}$ tensored with the Fock space$\mathcal{F}_{0}$
such that $W\left( t\right) \left( \mathrm{I}\otimes \delta _{\varnothing
}\right) =V\left( t\right) $, where $\delta _{\varnothing }\in \mathcal{F}%
_{0}$ is the Fock vacuum vector. In fact the state diffusion equation was
first derived \cite{Be88, Be89b} in this way from even more general quantum
stochastic unitary evolution which satisfy the equation 
\begin{equation*}
\left( \mathrm{I}\otimes \delta _{\varnothing }\right) ^{\dagger }W\left(
t\right) ^{\dagger }\left( \mathrm{B}\otimes g\left( \hat{w}_{0}^{t]}\right)
\right) W\left( t\right) \left( \mathrm{I}\otimes \delta _{\varnothing
}\right) =\mathsf{M}\left[ gV\left( t\right) ^{\dagger }\mathrm{B}V\left(
t\right) \right] .
\end{equation*}%
Indeed, this equation is satisfied for the model (\ref{6.4}) as one can
easily check for 
\begin{equation*}
g\left( w_{0}^{t]}\right) =\exp \left[ \int_{0}^{t}\left( f^{k}\left(
r\right) \mathrm{d}w_{k}^{r}-\frac{1}{2}f\left( r\right) ^{2}\mathrm{d}%
r\right) \right]
\end{equation*}%
given by a test vector function $f$ by conditioning the Langevin equation (%
\ref{6.5}) with respect to the vacuum vector $\delta _{\varnothing }$: 
\begin{equation*}
\left( \mathrm{I}\otimes \delta _{\varnothing }\right) ^{\dagger }\left( 
\mathrm{d}X+\left( K^{\dagger }X+XK-L^{\dagger }XL-\left( XL+L^{\dagger
}X\right) f\right) \mathrm{d}t\right) \left( \mathrm{I}\otimes \delta
_{\varnothing }\right) =0.
\end{equation*}%
Obviously this equation coincides with the conditional expectation 
\begin{equation*}
\mathsf{M}\left[ \mathrm{d}B\left( t\right) +\left( \mathrm{K}^{\dagger
}B\left( t\right) +B\left( t\right) \mathrm{K}-\mathrm{L}^{\dagger }B\left(
t\right) \mathrm{L}-\left( B\left( t\right) L+L^{\dagger }B\left( t\right)
\right) f\left( t\right) \right) \mathrm{d}t\right] =0
\end{equation*}%
for the stochastic process $B\left( t\right) =V\left( t\right) ^{\dagger }g%
\mathrm{X}V\left( t\right) $ which satisfies the stochastic It\^{o} equation 
\begin{equation*}
\mathrm{d}B+gV^{\dagger }\left( \mathrm{K}^{\dagger }\mathrm{X}+\mathrm{XK}-%
\mathrm{L}^{\dagger }\mathrm{XL}-\left( \mathrm{XL}+\mathrm{L}^{\dagger }%
\mathrm{X}\right) f\right) V\mathrm{d}t=gV^{\dagger }\left( \mathrm{L}%
^{\dagger }\mathrm{X}+\mathrm{XL}+\mathrm{X}\right) V\mathrm{d}w^{t}.
\end{equation*}%
This however doesn't give yet the complete solution of the quantum
measurement problem as formulated above because the algebra $\mathfrak{B}%
_{0} $ generated by $\mathrm{B}\otimes I$ and the Langevin forces $\hat{f}%
_{k}^{t} $ does not contain the measurement processes $\hat{w}_{k}^{t}$
which do not commute with $\hat{f}_{k}^{t}$, and the unitary family $W\left(
t\right) $ does not form unitary group but only cocycle 
\begin{equation*}
T_{t}W\left( s\right) T_{-t}W\left( t\right) =W\left( s+t\right) ,\quad
\forall s,t>0
\end{equation*}%
with respect to the isometric but not unitary right shift semigroup $T_{t}$
in $\mathcal{F}_{0}$.

Let $T_{t}$ be the one parametric continuous unitary shift group on $%
\mathcal{F}^{0]}\otimes \mathcal{F}_{0}$ extending the definition from $%
\mathcal{F}_{0}$. It describes the free evolution by right shifts $\Phi
_{t}\left( \omega \right) =\Phi \left( \omega -t\right) $ in Fock space over
the whole line $\mathbb{R}$. Then one can easily find the unitary group 
\begin{equation*}
U^{t}=T_{-t}\left( I^{0}\otimes \mathrm{I}\otimes W\left( t\right) \right)
T_{t}
\end{equation*}%
on $\mathcal{F}_{0}\otimes \mathfrak{h}\otimes \mathcal{F}^{0]}$ inducing
the quantum stochastic evolution as the interaction representation $U\left(
t\right) =T_{t}U^{t}$ on the Hilbert space $\mathfrak{h}\otimes \mathcal{G}%
_{0}$. In fact this evolution corresponds to an unphysical coordinate
discontinuity problem at the origin $s=0$ which is not invariant under the
reflection of time $t\mapsto -t$. Instead, we shall formulate the unitary
equivalent boundary value problem in the Poisson space $\mathcal{G}=\mathcal{%
G}_{-}\otimes \mathcal{G}_{+}$ for two semi-infinite strings on $\mathbb{R}%
_{+}$, one is the living place for the quantum noise generated by a Poisson
flow of incoming waves of quantum particles of the intensity $\nu >0$, and
the other one is for the outgoing classical particles carrying the
information after a unitary interaction with the measured quantum system at
the origin $r=0$. The probability amplitudes $\Phi \in \mathcal{G}$ are
represented by the $\mathbb{G}^{\otimes }=\mathfrak{g}_{-}^{\otimes }\otimes 
\mathfrak{g}_{+}^{\otimes }$-valued functions $\Phi \left( \upsilon
_{-},\upsilon _{+}\right) $ of two infinite sequences $\upsilon _{\pm
}=\left\{ \pm r_{1},\pm r_{2},\ldots \right\} \subset \mathbb{R}_{+}$ of the
coordinates of the particles in the increasing order $r_{1}<r_{2}<\ldots $
such that 
\begin{equation*}
\left\Vert \Phi \right\Vert ^{2}=\iint \left\Vert \Phi \left( \upsilon
_{-},\upsilon _{+}\right) \right\Vert ^{2}\mathsf{P}_{\nu }\left( \mathrm{d}%
\upsilon _{-}\right) \mathsf{P}_{\nu }\left( \mathrm{d}\upsilon _{+}\right)
<\infty
\end{equation*}%
with respect to the product of two copies of the Poisson probability measure 
$\mathsf{P}_{\nu }$ defined by the constant intensity $\nu >0$ on $\mathbb{R}%
_{+}$. Here $\mathfrak{g}^{\otimes }$ is the infinite tensor product of $%
\mathfrak{g}=\mathbb{C}^{d}$ obtained by the completion of the linear span
of $\chi _{1}\otimes \chi _{2}\otimes \ldots $ with almost all multipliers\ $%
\chi _{n}=\varphi $ given by a unit vector $\varphi \in \mathbb{C}^{d}$ such
that the infinite product $\left\Vert \Phi \left( \upsilon \right)
\right\Vert =\prod_{r\in \upsilon }\left\Vert f\left( r\right) \right\Vert $
for $\Phi \left( \upsilon \right) =\otimes _{r\in \upsilon }f\left( r\right) 
$ with $f\left( r_{n}\right) =\chi _{n}$ is well defined as it has all but
finite number of multipliers $\left\Vert \chi _{n}\right\Vert $ equal $1$.
The unitary transformation $\digamma \mapsto \Phi $ from a Fock space $%
\mathcal{F}\ni \digamma $ to the corresponding Poisson one $\mathcal{G}$ can
be written as 
\begin{equation*}
\Phi =\lim_{t\rightarrow \infty }\mathrm{e}^{\varphi ^{i}A_{i}^{+}\left(
t\right) }\mathrm{e}^{-\varphi _{k}A_{-}^{k}\left( t\right) }\nu ^{-\frac{1}{%
2}A_{i}^{i}\left( t\right) }\digamma \equiv I_{\nu }\left( \varphi \right)
\digamma ,
\end{equation*}%
where $\varphi _{k}=\nu \bar{\varphi}^{k}$ for the Poisson intensity $\nu >0$
and the unit vector $\varphi =\left( \varphi ^{i}\right) $ defined by the
initial probability amplitude $\varphi \in \mathfrak{g}$ for the auxiliary
particles to be in a state $k=1,\ldots ,d.$ Here $A_{\iota }^{\kappa }\left(
t\right) $ are the QS integrators defined in the Appendix, and the limit is
taken on the dense subspace $\cup _{t>0}\mathcal{F}_{0}^{t]}$ of
vacuum-adapted Fock functions $\digamma _{t}\in \mathcal{F}_{0}$ and
extended then onto $\mathcal{F}_{0}$ by easily proved isometry $\left\Vert
\digamma _{t}\right\Vert =\left\Vert \Phi _{t}\right\Vert $ for $\Phi
_{t}=I_{\nu }\left( \varphi \right) \digamma _{t}$.

The free evolution in $\mathcal{G}$ is the left shift for the incoming waves
and the right shift for outgoing waves, 
\begin{equation*}
T_{t}\Phi \left( \upsilon _{-},\upsilon _{+}\right) =\Phi \left( \upsilon
_{-}^{t},\upsilon _{+}^{t}\right) ,
\end{equation*}
where $\upsilon _{\pm }^{t}=\pm \left[ \left( \left[ \left( -\upsilon
_{-}\right) \cup \left( +\upsilon _{+}\right) \right] -t\right) \cap \mathbb{%
R}_{\pm }\right] $. It is given by the second quantization 
\begin{equation*}
P\Phi \left( \upsilon _{-},\upsilon _{+}\right) =\frac{\hbar }{\mathrm{i}}%
\left( \sum_{r\in \upsilon _{+}}\frac{\partial }{\partial r}-\sum_{r\in
\upsilon _{-}}\frac{\partial }{\partial r}\right) \Phi \left( \upsilon
_{-},\upsilon _{+}\right)
\end{equation*}
of the Dirac Hamiltonian in one dimension on $\mathbb{R}_{+}.$

In order to formulate the boundary value problem in the space $\mathcal{H}=%
\mathfrak{h}\otimes\mathcal{G}$ corresponding to the quantum stochastic
equations of the diffusive type (\ref{6.4}) let us introduce the notation 
\begin{equation*}
\Phi\left( 0^{k}\sqcup\upsilon_{\pm}\right) =\lim_{r\searrow0}\left( \langle
k|\otimes\mathrm{I}_{1}\otimes\mathrm{I}_{2}\ldots\right) \Phi\left( \pm
r,\pm r_{1},\pm r_{2},\ldots\right) ,
\end{equation*}
where $\langle k|=d^{-1/2}\left( \delta_{1}^{k},\ldots,\delta_{l}^{k}\right) 
$ acts as the unit bra-vector evaluating the $k$-th projection of the state
vector $\Phi\left( \pm r\sqcup\upsilon_{\pm}\right) $ with $%
r<r_{1}<r_{2}<\ldots$ corresponding to the nearest to the boundary $r=0$
particle in one of the strings on $\mathbb{R}_{+}$.

The unitary group evolution $U^{t}$ corresponding to the scattering
interaction at the boundary with the continuously measured system which has
its own free evolution described by the energy operator $\mathrm{E}=\mathrm{E%
}^{\dagger }$ can be obtained by resolving the following generalized Schr%
\"{o}dinger equation 
\begin{equation}
\frac{\partial }{\partial t}\Psi ^{t}\left( \upsilon _{-},\upsilon
_{+}\right) =\frac{\mathrm{i}}{\hbar }P\Psi ^{t}\left( t,\upsilon
_{-},\upsilon _{+}\right) +\mathrm{G}_{+}^{-}\Psi ^{t}\left( \upsilon
_{-},\upsilon _{+}\right) +\mathrm{G}_{k}^{-}\Psi ^{t}\left( \upsilon
_{-},0^{k}\sqcup \upsilon _{+}\right)  \label{6.7}
\end{equation}
with the Dirac zero current boundary condition at the origin $r=0$ 
\begin{equation}
\Psi ^{t}\left( 0^{i}\sqcup \upsilon _{-},\upsilon _{+}\right) =\mathrm{G}%
_{+}^{i}\Psi ^{t}\left( \upsilon _{-},\upsilon _{+}\right) +\mathrm{G}%
_{k}^{i}\Psi ^{t}\left( \upsilon _{-},0^{k}\sqcup \upsilon _{+}\right)
,\quad \forall t>0,\upsilon _{\pm }>0.  \label{6.8}
\end{equation}
Here $\mathrm{G}=\left[ \mathrm{G}_{k}^{i}\right] $ is unitary, $\mathrm{G}%
^{-1}=\mathrm{G}^{\dagger }$, the scattering operator $\mathrm{S}$ in the
simpler quantum jump boundary value problem corresponding to $\mathrm{G}%
_{+}=0=\mathrm{G}^{-}$, and the other system operators $\mathrm{G}_{\iota
}^{\kappa }$, with $\iota =-,i$ and $\kappa =k,+$ for any $i,k=1,\ldots ,d$
are chosen as 
\begin{equation}
\mathrm{G}^{-}+\nu \mathrm{G}_{+}^{\dagger }\mathrm{G}=\mathrm{O},\quad 
\mathrm{G}_{+}^{-}+\frac{\nu }{2}\mathrm{G}_{+}^{\dagger }\mathrm{G}_{+}+%
\frac{i}{\hbar }\mathrm{E}=\mathrm{O}.  \label{6.9}
\end{equation}
Note that these conditions can be written as pseudo-unitarity operator of
the following triangular block-matrix 
\begin{equation*}
\left[ 
\begin{tabular}{lll}
\textrm{I} & \textrm{G}$^{-}$ & \textrm{G}$_{+}^{-}$ \\ 
\textrm{O} & \textrm{G} & \textrm{G}$_{+}$ \\ 
\textrm{O} & \textrm{O} & \textrm{I}%
\end{tabular}
\right] ^{-1}=\left[ 
\begin{tabular}{lll}
\textrm{O} & \textrm{O} & \textrm{I} \\ 
\textrm{O} & $\nu $\textrm{I} & \textrm{O} \\ 
\textrm{I} & \textrm{O} & \textrm{O}%
\end{tabular}
\right] ^{-1}\left[ 
\begin{tabular}{lll}
\textrm{I} & \textrm{G}$^{-}$ & \textrm{G}$_{+}^{-}$ \\ 
\textrm{O} & \textrm{G} & \textrm{G}$_{+}$ \\ 
\textrm{O} & \textrm{O} & \textrm{I}%
\end{tabular}
\right] ^{\dagger }\left[ 
\begin{tabular}{lll}
\textrm{O} & \textrm{O} & \textrm{I} \\ 
\textrm{O} & $\nu $\textrm{I} & $\mathrm{O}$ \\ 
\textrm{1} & \textrm{O} & \textrm{O}%
\end{tabular}
\right] .
\end{equation*}
As it was proved in \cite{Be88a, Be92a} this is a necessary (and sufficient
if all operators are bounded) condition for the unitarity $W\left( t\right)
^{-1}=W\left( t\right) ^{\dagger }$ of the cocycle solution resolving the
quantum stochastic differential equation 
\begin{equation*}
\mathrm{d}\Psi _{0}\left( t\right) =\left( \mathrm{G}_{\kappa }^{\iota
}-\delta _{\kappa }^{\iota }\mathrm{I}\right) \Psi _{0}\left( t\right) 
\mathrm{d}A_{\iota }^{\kappa },\quad \Psi _{0}\left( 0\right) =\Psi _{0}
\end{equation*}
in the Hilbert space $\mathcal{H}_{0}=\mathfrak{h}\otimes \mathcal{G}_{0}$
where $\mathcal{G}_{0}$ is identified with the space $\mathcal{G}_{+}=%
\mathbb{G}^{\otimes }\otimes L_{\mu }^{2}$ for the Poisson measure $\mu =%
\mathsf{P}_{\nu }$ with the intensity $\nu $ on $\mathbb{R}_{+}$. This is
the general form for the quantum stochastic equation (\ref{6.4}) where $%
\mathrm{d}A_{-}^{+}=\mathrm{d}t$ in the Poisson space (see the Appendix for
more detail explanations of these notations). Our recent results partially
published in \cite{Be00, Be01a, BeKo01} prove that this quantum stochastic
evolution extended as the identity $I_{-}$ also on the component $\mathcal{G}%
_{-}$ for the scattered particles, is nothing but the interaction
representation $U^{t}=T_{-t}\left( I_{-}\otimes W\left( t\right) \right) $
for the unitary group $U^{t}$ resolving our boundary value problem in $%
\mathfrak{h}\otimes \mathbb{G}^{\otimes }$ times the Poisson space $L_{\mu
}^{2}$. Thus the pseudounitarity condition (\ref{6.9}) is necessary (and
sufficient if the operators $\mathrm{G}_{\kappa }^{\iota }$ are bounded) for
the self-adjointness of the Dirac type boundary value problem (\ref{6.7}), (%
\ref{6.8}).

The generators $\mathrm{G}_{\kappa}^{\iota}$ of this boundary value problem
define the generators $\mathrm{S}_{\kappa}^{\iota}$ of the corresponding
quantum stochastic equation in Fock space by the following transformation 
\begin{align}
\mathrm{S}_{+}^{i} & =\nu^{1/2}\left( \mathrm{G}_{+}^{i}+\mathrm{G}%
_{k}^{i}\varphi^{k}-\varphi^{i}\right) ,\;\mathrm{S}_{k}^{-}=\nu^{-1/2}%
\left( \mathrm{G}_{k}^{-}+\varphi_{i}\mathrm{G}_{k}^{i}-\varphi_{k}\right) 
\notag \\
\mathrm{S}_{+}^{-} & =\mathrm{G}_{+}^{-}+\varphi_{i}\mathrm{G}_{+}^{i}+%
\mathrm{G}_{k}^{-}\varphi^{k}+\varphi_{i}\left( \mathrm{G}%
_{k}^{i}-\delta_{k}^{i}\mathrm{I}\right) \varphi^{k},\quad\quad\mathrm{S}%
_{k}^{i}=\mathrm{G}_{k}^{i},\;
\end{align}
induced by the canonical transformation $I_{\nu}\left( \varphi\right) $.

The quantum state diffusion equation (\ref{6.6}) for the continuous
measurement of the coordinates $\mathrm{Q}^{k}$ corresponds to the
particular case (\ref{6.2})of the quantum stochastic differential equation
in Fock space, with 
\begin{align*}
\mathrm{S}_{+}^{i} & =\nu^{1/2}\mathrm{G}_{+}^{i},\quad\quad\quad \quad%
\mathrm{S}_{k}^{-}=\nu^{-1/2}\mathrm{G}_{k}^{-} \\
\mathrm{S}_{+}^{-} & =\mathrm{G}_{+}^{-}+\varphi_{i}\mathrm{G}_{+}^{i}+%
\mathrm{G}_{k}^{-}\varphi^{k},\quad\mathrm{S}_{k}^{i}=\delta_{k}^{i}\mathrm{I%
},
\end{align*}
$\ $and $\mathrm{G}_{+}^{i}=\mathrm{Q}^{i}$, $\mathrm{G}_{k}^{-}=\nu \mathrm{%
Q}^{k}$ such that all coupling constants $\lambda_{k}=\nu^{1/2}$. are equal
to the square root of the flow intensity $\nu$. The operators $\mathrm{G}%
_{+}^{i}=\varphi^{i}\mathrm{Q}^{i}$ and $\mathrm{G}_{k}^{-}=\mathrm{Q}%
_{k}\varphi_{k}$ corresponding to the different \ couplings $\lambda_{k}$
can also be obtained from the purely jump model in the central limit $%
\nu\mapsto\infty$ as it was done in \cite{BeMe96}. In this case 
\begin{equation*}
\mathrm{S}_{+}^{i}=\nu^{1/2}\left( \mathrm{G}-\mathrm{I}\right)
_{k}^{i}\varphi^{k}\rightarrow-\mathrm{i}\lambda\varphi^{i}\mathrm{Q}^{i},
\end{equation*}
with $\varphi^{k}=\mathrm{i}\lambda_{k}/\lambda$.

And finally, we have to find the operator processes $Y_{k}^{s},s\leq0$ on
the Hilbert space $\mathcal{G}_{-}$ which reproduce the standard Wiener
noises $w_{k}^{t}$ in the state diffusion when our dynamical model is
conditioned (filtered) with respect to their nondemolition measurement. As
the candidates let us consider the field coordinate processes 
\begin{equation*}
X_{k}^{-t}=A_{k}^{+}(-t,0]+A_{-}^{k}(-t,0]=T_{-t}\left(
A_{k}^{+}(0,t]+A_{-}^{k}(0,t]\right) T_{t}
\end{equation*}
which are given by the creation and annihilation processes $A^{+}\left(
t\right) $ and $A_{-}\left( t\right) $ shifted from $\mathcal{G}_{+}$. In
our Poisson space model of $\mathcal{G}$ they have not zero expectations 
\begin{equation*}
\Phi^{\dagger}X_{k}^{-t}\Phi=\Phi^{\dagger}\left(
A_{k}^{+}(0,t]+A_{-}^{k}(0,t]\right) \Phi=2\nu^{1/2}t
\end{equation*}
in the ground state $\Phi=I_{\nu}\left( \varphi\right) \delta_{\varnothing}$
corresponding to the vacuum vector $\delta_{\varnothing}$ in the Fock space.
This state is given as the infinite tensor product $\Phi^{\circ}=\varphi
_{-}^{\otimes}\otimes\varphi_{+}^{\otimes}$ of all equal probability
amplitudes $\varphi_{-}=\varphi=\varphi_{+}$ in $\mathfrak{g}=\mathbb{C}^{d}$
for each sequence $\upsilon_{-}$ and $\upsilon_{+}$. Hence the independent
increment processes $Y_{k}^{t}=T_{t}Y_{k}^{-t}T_{-t}$ corresponding to the
standard Wiener noises $w_{k}^{t}$ represented in Fock spaces as $\hat{w}%
_{k}^{t}=A_{k}^{+}\left( t\right) +A_{-}^{k}\left( t\right) $ are the
compensated processes $Y_{k}^{-t}=X_{k}^{-t}-2\nu^{1/2}t$. This unitary
equivalence of $Y_{k}^{t}$ and $\hat{w}_{k}^{t}$ under the Fock-Poisson
transformation $I_{\nu}\left( \varphi\right) $, and the deduction given
above of the quantum state diffusion from the quantum stochastic signal plus
noise model (\ref{6.1}) for continuous observation in Fock space, completes
the solution of the quantum measurement model in its rigorous formulation.

\section{Conclusion: A quantum message from the future}

Recent phenomenological theories of continuous reduction, quantum state
diffusion and quantum trajectories extended the instantaneous projection
postulate to a certain class of continuous-in-time measurements. As was
shown here, there is no need to supplement the usual quantum mechanics with
any of such generalized reduction postulate even in the continuous time.
They all have been derived from the time continuous unitary evolution for a
generalized Dirac type Schr\"{o}dinger equation with a singular scattering
interaction at the boundary of our Hamiltonian model, see the recent review
paper \cite{Be02}. The quantum causality as a new superselection rule
provides a time continuous nondemolition measurement in the extended system
which enables to obtain the quantum state diffusion and quantum trajectories
simply by time continuous conditioning called quantum filtering. Our
nondemolition causality principle, which was explicitly formulated as a
causal commutativity condition in \cite{Be79, Be80}, admits to select a
continuous diffusive classical process in the quantum extended world which
satisfies the nondemolition condition with respect to all future of the
measured system. And this allows\ us to obtain the continuous trajectories
for quantum state diffusion by simple filtering of quantum noise exactly as
it was done in the classical statistical nonlinear filtering and prediction
theory. In this way we derived the quantum state diffusion of a Gaussian
wave packet already in \cite{Be79, Be80} as the result of the solution of
quantum prediction problem by filtering the quantum white noise in a quantum
stochastic Langevin model for the continuous observation. Thus the
\textquotedblleft primary\textquotedblright\ for the conventional quantum
mechanics stochastic nonlinear irreversible quantum state diffusion appears
to be the secondary, as it should be, to the deterministic linear unitary
reversible evolution of the extended quantum mechanics containing
necessarily infinite number of auxiliary particles. However quantum
causality, which defines the arrow of time by selecting what part of the
reversible world is related to the classical past and what is related to the
quantum future, makes the extended mechanics irreversible in terms of the
injective semigroup of the invertible Heisenberg transformations induced by
the unitary group evolution for the positive arrow of time. The microscopic
information dynamics of this event enhanced quantum mechanics, or Eventum
Mechanics, allows the emergence of the decoherence and the increase of
entropy in a purely dynamical way without any sort of reservoir averaging.

Summarizing, we can formulate the general principles of the Eventum
Mechanics which unifies the classical and quantum mechanics in such a way
that there is no contradiction between the unitary evolution of the matter
waves and the phenomenological information dynamics such as quantum state
diffusion or spontaneous jumps for the events and the trajectories of the
particles. This can be described in the Schr\"{o}dinger picture as an
extended, \emph{non-stochastic} unitary evolution quantum mechanics with
causality as a superselection rule making it time asymmetric in the
Heisenberg picture where it is described by endomorphic irreversible quantum
dynamics in the following way:

\begin{itemize}
\item It is a reversible wave mechanics of the continuous unitary group
evolutions in an infinite-dimensional Hilbert space

\item It has conventional interpretation for the normalized Hilbert space
vectors\ as state-vectors ( probability amplitudes)

\item However not all operators, e.g. the dynamical generator (Hamiltonian),
are admissible as the potential observables

\item Quantum causality is statistical predictability of the quantum states
based on the results of the actual measurements

\item It implies the choice of time arrow and an initial state which
together with past measurement data defines the reality

\item The actual observables (beables) must be compatible with any operator
representing a potential (future) observable

\item The Heisenberg dynamics and others symmetries induced by unitary
operators should be algebraically endomorphic

\item However these endomorphisms form only a semigroup on the algebra of
all observables as they may be irreversible.
\end{itemize}

Note that the classical Hamiltonian mechanics can be also described in this
way by considering only the commutative algebras of the potential
observables. Each such observable is compatible with any other and can be
considered as an actual observable, or beable. However, the Hamiltonian
operator, generating a non-trivial Liouville unitary dynamics in the
corresponding Hilbert space, is not an observable, as it doesn't commute
with any observable which is not the integral of motion. Nevertheless the
corresponding Heisenberg dynamics, described by the induced automorphisms of
the commutative algebra, is reversible, and pure states, describing the
reality, remain pure, non disturbed by the measurements of its observables.
This is also true in the purely quantum mechanical case, in which the
Hamiltonian is an observable, as there are no events and nontrivial beables
in the conventional quantum mechanics. The only actual observables, which
are compatible with any \ Hermitian operator as a potential observable, are
the constants, i.e. proportional to the identity operator, as the only
operators, commuting with any such observable. Their measurements do not
bring new information\ and do not disturb the quantum states. However any
non-trivial classical--quantum Hamiltonian interactions cannot induce a
group of the reversible Heisenberg automorphisms but only a semigroup of
irreversible endomorphisms of the decomposable algebra of all potential
observables of the composed classical-quantum system. This follows from the
simple fact that any automorphism leaves the center of an operator algebra
invariant, and thus induces the autonomous noninteracting dynamics on the
classical part of the semi-classical system. This is the only reason which
is responsible for failure of all earlier desperate attempts to build the
reversible, time symmetric Hamiltonian theory of classical-quantum
interaction which would give a dynamical solution of the quantum decoherence
and measurement problem along the line suggested by von Neumann and Bohr.
There is no nontrivial reversible classical-quantum mechanical interaction,
but as we have seen, there is a Hamiltonian irreversible interaction within
the time asymmetric Eventum Mechanics..

The unitary solution of the described boundary value problem indeed induces
endomorphic semi-classical Hamiltonian dynamics, and in fact is underlying
in any phenomenological reduction model \cite{Be02}. Note that although the
irreversible Heisenberg endomorphisms of eventum mechanics, induced by the
unitary propagators, are injective, and thus are invertible by completely
positive maps, and are not mixed, they mix the pure states over the center
of the algebra. Such mixed states, which are uniquely represented as the
orthogonal mixture over the `hidden' variables (beables), can be filtered by
the measurement of the actual observables, and this transition from the
prior state corresponding to the less definite (mixed) reality to the
posterior state, corresponding to a more definite (pure) reality by the
simple inference is not change the reality. This is an explanation, in the
pure dynamical terms of the eventum mechanics, of the emergence of the
decoherence and the reductions due to the measurement, which has no
explanation in the conventional classical and quantum mechanics.

Our mathematical formulation of the eventum mechanics as the extended
quantum mechanics equipped with the quantum causality to allow events and
trajectories in the theory, is just as continuous as Schr\"{o}dinger could
have wished. However, it doesn't exclude the jumps which only appear in the
singular interaction picture, which are there as a part of the theory, not
only of its interpretation. Although Schr\"{o}dinger himself didn't believe
in quantum jumps, he tried several times, although unsuccessfully, to obtain
the continuous reduction from a generalized, relativistic, ``true Schr\"{o}%
dinger'' equation. He envisaged that `if one introduces two symmetric
systems of waves, which are traveling in opposite directions; one of them
presumably has something to do with the known (or supposed to be known)
state of the system at a later point in time' \cite{Schr31}, then it would
be possible to derive the `verdammte Quantenspringerei' for the opposite
wave as a solution of the future-past boundary value problem. This desire
coincides with the ``transactional'' \ attempt of interpretation of quantum
mechanics suggested in \cite{Crm86} on the basis that the relativistic wave
equation yields in the nonrelativistic limit two Schr\"{o}dinger type
equations, one of which is the time reversed version of the usual equation:
`The state vector $\psi $ of the quantum mechanical formalism is a real
physical wave with spatial extension and it is identical with the initial
``offer wave'' of the transaction. The particle (photon, electron, etc.) and
the collapsed state vector are identical with the completed transaction.' \
There was no proof of this conjecture, and now we know that it is not even
possible to derive the quantum state diffusions, spontaneous jumps and
single reductions from models involving only a finite particle state vectors 
$\psi \left( t\right) $ satisfying the conventional Schr\"{o}dinger equation.

Our new approach, based on the exactly solvable boundary value problems for
infinite particle states described in this paper, resolves the problem
formulated by Schr\"{o}dinger. And thus it resolves the old problem of
interpretation of the quantum theory, together with its infamous paradoxes,
in a constructive way by giving exact nontrivial models for allowing the
mathematical analysis of quantum observation processes determining the
phenomenological coupling constants and the reality underlying these
paradoxes. Conceptually it is based upon a new idea of quantum causality
called the nondemolition principle \cite{Be94} which divides the world into
the classical past, forming the consistent histories, and the quantum
future, the state of which is predictable for each such history.

\section{\textsc{Appendix}}

\subsection{Symbolic Quantum Calculus and Stochastic Differential Equations.}

In order to formulate the differential nondemolition causality condition and
to derive a filtering equation for the posterior states in the
time-continuous case we need quantum stochastic calculus.

The classical differential calculus for the infinitesimal increments 
\begin{equation*}
\mathrm{d}x=x\left( t+\mathrm{d}t\right) -x\left( t\right)
\end{equation*}
became generally accepted only after Newton gave a simple algebraic rule $%
\left( \mathrm{d}t\right) ^{2}=0$ for the formal computations of the
differentials $\mathrm{d}x$ for smooth trajectories $t\mapsto x\left(
t\right) $. In the complex plane $\mathbb{C}$ of phase space it can be
represented by a one-dimensional algebra $\mathfrak{\alpha}=\mathbb{C}%
\mathrm{d}_{t}$ of the elements $a=\alpha\mathrm{d}_{t}$ with involution $%
a^{\star}=\bar{\alpha}\mathrm{d}_{t}$. Here 
\begin{equation*}
\text{$\mathrm{d}_{t}$}=\left[ 
\begin{array}{ll}
0 & 1 \\ 
0 & 0%
\end{array}
\right] =\frac{1}{2}\left( \sigma_{x}+i\sigma_{y}\right)
\end{equation*}
for $\mathrm{d}t$ is the nilpotent matrix, which can be regarded as
Hermitian $\mathrm{d}_{t}^{\star}=\mathrm{d}_{t}$ with respect to the
Minkowski metrics $\left( \mathbf{z}|\mathbf{z}\right) =2\func{Re}z_{-}\bar{z%
}_{+}$ in $\mathbb{C}^{2}$.

This formal rule was generalized to non-smooth paths early in the last
century in order to include the calculus of forward differentials $\mathrm{d}%
w\simeq\left( \mathrm{d}t\right) ^{1/2}$ for continuous diffusions $w_{t}$
which have no derivative at any $t$, and the forward differentials $\mathrm{d%
}n\in\left\{ 0,1\right\} $ for left continuous counting trajectories $n_{t}$
which have zero derivative for almost all $t$ (except the points of
discontinuity where $\mathrm{d}n=1$). The first is usually done by adding
the rules 
\begin{equation*}
\left( \mathrm{d}w\right) ^{2}=\mathrm{d}t,\quad\mathrm{d}w\mathrm{d}t=0=%
\mathrm{d}t\mathrm{d}w
\end{equation*}
in formal computations of continuous trajectories having the first order
forward differentials $\mathrm{d}x=\alpha\mathrm{d}t+\beta\mathrm{d}w$ with
the diffusive part given by the increments of standard Brownian paths $w $.
The second can be done by adding the rules 
\begin{equation*}
\left( \mathrm{d}n\right) ^{2}=\mathrm{d}n,\quad\mathrm{d}n\mathrm{d}t=0=%
\mathrm{d}t\mathrm{d}n
\end{equation*}
in formal computations of left continuous and smooth for almost all $t$
trajectories having the forward differentials $\mathrm{d}x=\alpha \mathrm{d}%
t+\gamma\mathrm{d}m$ with jumping part given by the increments of standard
compensated Poisson paths $m_{t}=n_{t}-t$. These rules were developed by It%
\^{o} \cite{Ito51} into the form of a stochastic calculus.

The linear span of $\mathrm{d}t$ and $\mathrm{d}w$ forms the Wiener-It\^{o}
algebra $\mathfrak{b}=\mathbb{C}\mathrm{d}_{t}+\mathbb{C}\mathrm{d}_{w}$,
while the linear span of $\mathrm{d}t$ and $\mathrm{d}n$ forms the Poisson-It%
\^{o} algebra $\mathfrak{c}=\mathbb{C}\mathrm{d}_{t}+\mathbb{C}\mathrm{d}%
_{m} $, with the second order nilpotent $\mathrm{d}_{w}=\mathrm{d}%
_{w}^{\star}$ and the idempotent $\mathrm{d}_{m}=\mathrm{d}_{m}^{\star}$.
They are represented together with $\mathrm{d}_{t}$ by the triangular
Hermitian matrices 
\begin{equation*}
\text{$\mathrm{d}_{t}$}=\left[ 
\begin{array}{lll}
0 & 0 & 1 \\ 
0 & 0 & 0 \\ 
0 & 0 & 0%
\end{array}
\right] ,\quad\mathrm{d}_{w}=\left[ 
\begin{array}{lll}
0 & 1 & 0 \\ 
0 & 0 & 1 \\ 
0 & 0 & 0%
\end{array}
\right] ,\emph{\quad}\mathrm{d}_{m}\mathbf{=}\left[ 
\begin{array}{lll}
0 & 1 & 0 \\ 
0 & 1 & 1 \\ 
0 & 0 & 0%
\end{array}
\right] ,
\end{equation*}
on the Minkowski space $\mathbb{C}^{3}$ with respect to the inner Minkowski
product $\left( \mathbf{z}|\mathbf{z}\right) =z_{-}z^{-}+z_{\circ}z^{\circ
}+z_{+}z^{+}$, where $z^{\mu}=\bar{z}_{-\mu}$, $-\left( -,\circ,+\right)
=\left( +,\circ,-\right) $.

Although both algebras $\mathfrak{b}$ and $\mathfrak{c}$ are commutative,
the matrix algebra $\mathfrak{a}$ generated by $\mathfrak{b}$ and $\mathfrak{%
c}$ on $\mathbb{C}^{3}$ is not: 
\begin{equation*}
\mathrm{d}_{w}\mathrm{d}_{m}=\left[ 
\begin{array}{lll}
0 & 1 & 1 \\ 
0 & 0 & 0 \\ 
0 & 0 & 0%
\end{array}
\right] \neq\left[ 
\begin{array}{lll}
0 & 0 & 1 \\ 
0 & 0 & 1 \\ 
0 & 0 & 0%
\end{array}
\right] =\mathrm{d}_{m}\mathrm{d}_{w}.
\end{equation*}
The four-dimensional $\star$-algebra $\mathfrak{a}=\mathbb{C}\mathrm{d}_{t}+%
\mathbb{C}\mathrm{d}_{-}+\mathbb{C}\mathrm{d}^{+}+\mathbb{C}\mathrm{d}$ of
triangular matrices with the canonical basis 
\begin{equation*}
\mathrm{d}_{-}=\left[ 
\begin{array}{lll}
0 & 1 & 0 \\ 
0 & 0 & 0 \\ 
0 & 0 & 0%
\end{array}
\right] ,\,\mathrm{d}^{+}\mathbf{=}\left[ 
\begin{array}{lll}
0 & 0 & 0 \\ 
0 & 0 & 1 \\ 
0 & 0 & 0%
\end{array}
\right] ,\,\mathrm{d}=\left[ 
\begin{array}{lll}
0 & 0 & 0 \\ 
0 & 1 & 0 \\ 
0 & 0 & 0%
\end{array}
\right] ,
\end{equation*}
given by the algebraic combinations 
\begin{equation*}
\mathrm{d}_{-}=\mathrm{d}_{w}\mathrm{d}_{m}-\text{$\mathrm{d}_{t}$},\;%
\mathrm{d}^{+}=\mathrm{d}_{m}\mathrm{d}_{w}-\text{$\mathrm{d}_{t}$},\;\text{%
\textrm{d}}=\text{\textrm{d}}_{m}-\text{\textrm{d}}_{w}
\end{equation*}
is the canonical representation of the differential $\star$-algebra for
one-dimensional vacuum noise in the unified quantum stochastic calculus \cite%
{Be88a, Be92a}. It realizes the HP (Hudson-Parthasarathy) table \cite{HuPa84}
\begin{equation*}
\mathrm{d}A_{-}\mathrm{d}A^{+}=\mathrm{d}t,\quad\mathrm{d}A_{-}\mathrm{d}A=%
\mathrm{d}A_{-},\quad\mathrm{d}A\mathrm{d}A^{+}=\text{\textrm{d}}%
A^{+},\quad\left( \mathrm{d}A\right) ^{2}=\mathrm{d}A,
\end{equation*}
with products equal zero for all other pairs, for the multiplication of the
canonical counting $\mathrm{d}A=\lambda\left( \mathrm{d}\right) $, creation $%
\mathrm{d}A^{+}=\lambda\left( \mathrm{d}^{+}\right) $, annihilation $\mathrm{%
d}A_{-}=\lambda\left( \mathrm{d}_{-}\right) $, and preservation $\mathrm{d}%
t=\lambda\left( \text{$\mathrm{d}_{t}$}\right) $ quantum stochastic
integrators in Fock space over $L^{2}\left( \mathbb{R}_{+}\right) $. As was
proved recently in \cite{Be98}, any generalized It\^{o} algebra describing a
quantum noise can be represented in the canonical way as a $\star$%
-subalgebra of a quantum vacuum algebra 
\begin{equation*}
\mathrm{d}A_{\mu}^{\kappa}\mathrm{d}A_{\iota}^{\nu}=\delta_{\iota}^{\kappa }%
\mathrm{d}A_{\mu}^{\nu},\quad\iota,\mu\in\left\{ -,1,\ldots,d\right\}
;\;\kappa,\nu\in\left\{ 1,\ldots,d,+\right\} ,
\end{equation*}
in the Fock space with several degrees of freedom $d$, where $\mathrm{d}%
A_{-}^{+}=\mathrm{d}t$ and $d$ is restricted by the doubled dimensionality
of quantum noise (could be infinite), similar to the representation of every
semi-classical system with a given state as a subsystem of quantum system
with a pure state. Note that in this quantum It\^{o} product formula $\delta
_{\kappa}^{\iota}=0$ if $\iota=+$ or $\kappa=-$ as $\delta_{\kappa}^{\iota
}\neq0$ only when $\iota=\kappa$.

The quantum It\^{o} product gives an explicit form 
\begin{equation*}
\mathrm{d}\psi\psi^{\dagger}+\psi\mathrm{d}\psi^{\dagger}+\mathrm{d}\psi%
\mathrm{d}\psi^{\dagger}=\left( \alpha_{\kappa}^{\iota}\psi^{\dagger
}+\psi\alpha_{\kappa}^{\star\iota}+\alpha_{j}^{\iota}a_{\kappa}^{\star
j}\right) _{\kappa}^{\iota}\mathrm{d}A_{\iota}^{\kappa}
\end{equation*}
of the term $\mathrm{d}\psi\mathrm{d}\psi^{\dagger}$ for the adjoint quantum
stochastic differentials 
\begin{equation*}
\mathrm{d}\psi=\alpha_{\kappa}^{\iota}\mathrm{d}A_{\iota}^{\kappa},\quad%
\mathrm{d}\psi^{\dagger}=\alpha_{\kappa}^{\star\iota}\mathrm{d}A_{\iota
}^{\kappa},
\end{equation*}
for evaluation of the product differential 
\begin{equation*}
\mathrm{d}\left( \psi\psi^{\dagger}\right) =\left( \psi+\mathrm{d}%
\psi\right) \left( \psi+\mathrm{d}\psi\right) ^{\dagger}-\psi\psi^{\dagger }.
\end{equation*}
Here $\alpha_{-\kappa}^{\star\iota}=\alpha_{-\iota}^{\kappa\dagger}$ is the
quantum It\^{o} involution with respect to the switch $-\left( -,+\right)
=\left( +,-\right) $, $-\left( 1,\ldots,d\right) =\left( 1,\ldots ,d\right) $%
, introduced in \cite{Be88a}, and the Einstein summation is always
understood over $\kappa=1,\ldots,d,+$; $\iota=-,1,\ldots,d$ and $k=1,\ldots
,d$. This is the universal It\^{o} product formula which lies in the heart
of the general quantum stochastic calculus \cite{Be88a, Be92a} unifying the
It\^{o} classical stochastic calculi with respect to the Wiener and Poisson
noises and the quantum differential calculi \cite{HuPa84, GaCo85} based on
the particular types of quantum It\^{o} algebras for the vacuum or finite
temperature noises. It was also extended to the form of quantum functional It%
\^{o} formula and even for the quantum nonadapted case in \cite{Be91, Be93}.

Every stationary classical (real or complex) process $x^{t}$, $t>0$ with $%
x^{0}=0$ and independent increments $x^{t+\Delta}-x^{t}$ has mean values $%
\mathsf{M}\left[ x^{t}\right] =\lambda t$. The compensated process $%
y^{t}=x^{t}-\lambda t$, which is called noise, has an operator
representation $\hat{x}^{t}$ in Fock space $\mathcal{F}_{0}$ the Hilbert
space $L^{2}\left( \mathbb{R}_{+}\right) $ in the form of the integral with
respect to basic processes $A_{j}^{+},A_{-}^{j},A_{k}^{i}$\ such that $%
\digamma=f\left( \hat{x}\right) \delta_{\varnothing}\simeq f\left( x\right) $%
\ in terms of the $L_{\mu}^{2}$ -- Fock isomorphism $f\leftrightarrow%
\digamma $ of the chaos expansions 
\begin{equation*}
f\left( x\right)
=\sum_{n=0}^{\infty}\idotsint_{0<r_{1}<\ldots<r_{n}}\digamma\left(
r_{1},\ldots r_{n}\right) \mathrm{d}y^{r_{1}}\cdots \mathrm{d}%
y^{r_{n}}\equiv\int\digamma\left( \upsilon\right) \mathrm{d}y^{\upsilon}
\end{equation*}
of the stochastic functionals $f\in L_{\mu}^{2}$ having the finite second
moments $\mathsf{M}\left[ \left| f\right| ^{2}\right] =\left\|
\digamma\right\| ^{2}$ and the Fock vectors $\digamma\in\mathcal{F}_{0}$.
The expectations of the Fock operators $f\left( \hat{x}\right) $ given by
the iterated stochastic integrals $f$ coincides on the vacuum state-vector $%
\delta_{\varnothing}\in\mathcal{F}_{0}$ with their expectation given by the
probability measure $\mu$: 
\begin{equation*}
\mathsf{M}\left[ f\left( x\right) \right] =\langle\delta_{\varnothing
}|f\left( \hat{x}\right) \delta_{\varnothing}\rangle=\digamma\left(
\varnothing\right) .
\end{equation*}
If its differential increments $\mathrm{d}x^{t}$ form a two dimensional It%
\^{o} algebra, $\hat{x}^{t}$ can be represented in the form of a commutative
combination of the three basic quantum stochastic increments $%
A=A_{0}^{0},A_{-}=A_{-}^{0},A^{+}=A_{0}^{+}$. The It\^{o} formula for the
process $x^{t}\;$given by the quantum stochastic differential 
\begin{equation*}
\mathrm{d}\hat{x}^{t}=\alpha\mathrm{d}A+\alpha^{-}\mathrm{d}A_{-}+\alpha _{+}%
\mathrm{d}A^{+}\mathrm{d}\psi+\alpha_{+}^{-}\mathrm{d}t
\end{equation*}
can be obtained from the HP product \cite{HuPa84} 
\begin{equation*}
\mathrm{d}\hat{x}^{t}\mathrm{d}\hat{x}^{t\dagger}=\alpha\alpha^{\dagger }%
\mathrm{d}A+\alpha^{-}\alpha^{\dagger}\mathrm{d}A_{-}+\alpha\alpha^{-\dagger
}\mathrm{d}A^{+}+\alpha^{-}\alpha^{-\dagger}\mathrm{d}t.
\end{equation*}

The noises $y_{k}^{t}=x_{k}^{t}-\lambda_{k}t$ with stationary independent
increments are called standard if they have the standard variance $\mathsf{M}%
\left[ \left( x^{t}\right) ^{2}\right] =t$. In this case 
\begin{equation*}
\hat{y}_{k}^{t}=\left( A_{k}^{+}+A_{-}^{k}+\varepsilon_{k}A_{k}^{k}\right)
\left( t\right) =\varepsilon_{k}m_{k}^{t}+\left( 1-\varepsilon_{k}\right)
w_{k}^{t},
\end{equation*}
where $\varepsilon_{k}\geq0$ is defined by the equation $\left( \mathrm{d}%
x_{k}^{t}\right) ^{2}-\mathrm{d}t=\varepsilon\mathrm{d}x_{k}^{t}$. Such, and
indeed higher dimensional, quantum noises for continuous measurements in
quantum optics were considered in \cite{GPZ92, DPZG92}.

The general form of a quantum stochastic decoherence equation, based on the
canonical representation of the arbitrary It\^{o} algebra for a quantum
noise in the vacuum of $d$ degrees of freedom, can be written as 
\begin{equation*}
\mathrm{d}\hat{\psi}\left( t\right) =\left( \mathrm{S}_{\kappa}^{\iota
}-\delta_{\kappa}^{\iota}\mathrm{I}\right) \mathrm{d}A_{\iota}^{\kappa}\hat{%
\psi}\left( t\right) ,\quad\hat{\psi}\left( 0\right) =\psi
\otimes\delta_{\varnothing},\;\psi\in\mathfrak{h}.
\end{equation*}
Here $\mathrm{L}_{\kappa}^{\iota}$ are the operators in the system Hilbert
space $\mathfrak{h}$ $\ni\psi$ with $\mathrm{S}_{\kappa}^{\star-}\mathrm{S}%
_{+}^{\kappa}=0$ for the mean square normalization 
\begin{equation*}
\hat{\psi}\left( t\right) ^{\dagger}\hat{\psi}\left( t\right) =\mathsf{M}%
\left[ \psi\left( t,\cdot\right) ^{\dagger}\psi\left( t,\cdot\right) \right]
=\psi^{\dagger}\psi
\end{equation*}
with respect to the vacuum of Fock space of the quantum noise, where the
Einstein summation is understood over all $\kappa=-,1,\ldots,d,+$ with the
agreement 
\begin{equation*}
\mathrm{S}_{-}^{-}=\mathrm{I}=\mathrm{S}_{+}^{+},\quad\,\mathrm{S}_{-}^{j}=%
\mathrm{O}=\mathrm{S}_{j}^{+},\quad j=1,\ldots,d
\end{equation*}
and $\delta_{\kappa}^{\iota}=1$ for all coinciding $\iota,\kappa\in\left\{
-,1,\ldots,d,+\right\} $ such that $\mathrm{L}_{\kappa}^{\iota}=\mathrm{S}%
_{\kappa}^{\iota}-\delta_{\kappa}^{\iota}\mathrm{I}=0$ whenever $\iota=+$ or 
$\kappa=-$. In the notations $\mathrm{S}_{+}^{j}=\mathrm{L}^{j}$, $\mathrm{S}%
_{+}^{-}=-\mathrm{K}$, $\mathrm{S}_{j}^{-}=-\mathrm{K}_{j}$, $j=1,\ldots,d$
the decoherence wave equation takes the standard form \cite{Be95, Be97} 
\begin{equation*}
\mathrm{d}\hat{\psi}\left( t\right) +\left( \mathrm{Kd}t+\mathrm{K}_{j}%
\mathrm{d}A_{-}^{j}\right) \hat{\psi}\left( t\right) =\left( \mathrm{L}^{j}%
\mathrm{d}A_{j}^{+}+\left( \mathrm{S}_{k}^{i}-\delta_{k}^{i}\mathrm{I}%
\right) \mathrm{d}A_{i}^{k}\right) \hat{\psi}\left( t\right) ,
\end{equation*}
where $A_{j}^{+}\left( t\right) ,A_{-}^{j}\left( t\right) ,A_{i}^{k}\left(
t\right) $ are respectively the canonical creation, annihilation and
exchange processes in Fock space, and the normalization condition is written
as $\mathrm{L}_{k}\mathrm{L}^{k}=\mathrm{K}+\mathrm{K}^{\dagger}$ with $%
\mathrm{L}_{k}^{\dagger}=\mathrm{L}^{k}$ (the Einstein summation is over $%
i,j,k=1,\ldots,d$).

Using the quantum It\^{o} formula one can obtain the corresponding equation
for the quantum stochastic density operator $\hat{\varrho}%
=\psi\psi^{\dagger} $ which is the particular case $\kappa=-,1,\ldots,d,+$
of the general quantum stochastic Master equation 
\begin{equation*}
\mathrm{d}\hat{\varrho}\left( t\right) =\left( \mathrm{S}_{\gamma}^{\iota }%
\hat{\varrho}\left( t\right) \mathrm{S}_{\kappa}^{\star\gamma}-\hat {\varrho}%
\left( t\right) \delta_{\kappa}^{\iota}\right) \mathrm{d}A_{\iota
}^{\kappa},\quad\hat{\varrho}\left( 0\right) =\rho,
\end{equation*}
where the summation over $\kappa=-,k,+$ is extended to infinite number of $%
k=1,2,\ldots$. This general form of the decoherence equation with $\mathrm{L}%
_{\kappa}^{\star-}\mathrm{L}_{+}^{\kappa}=\mathrm{O}$ corresponding to the
normalization condition $\left\langle \hat{\varrho}\left( t\right)
\right\rangle =\mathrm{Tr}\rho$ in the vacuum mean, was recently derived in
terms of quantum stochastic completely positive maps in \cite{Be95, Be97}.
Denoting $\mathrm{L}_{\kappa}^{-}=-\mathrm{K}_{\kappa}$, $\mathrm{L}%
_{+}^{\star\iota}=-\mathrm{K}^{\iota}$ such that $\mathrm{K}%
_{\iota}^{\dagger }=\mathrm{K}^{\iota}$, this can be written as 
\begin{equation*}
\mathrm{d}\hat{\varrho}\left( t\right) +\mathrm{K}_{\kappa}\hat{\varrho }%
\left( t\right) \mathrm{d}A_{-}^{\kappa}+\hat{\varrho}\left( t\right) 
\mathrm{K}^{\iota}\mathrm{d}A_{\iota}^{+}=\left( \mathrm{L}_{\kappa}^{j}\hat{%
\varrho}\left( t\right) \mathrm{L}_{j}^{\star\iota}-\hat{\varrho }\left(
t\right) \delta_{\kappa}^{\iota}\right) \mathrm{d}A_{\iota}^{\kappa},
\end{equation*}
or in the notation above, $\mathrm{K}_{+}=\mathrm{K},\mathrm{K}^{-}=\mathrm{K%
}^{\dagger}$, $\mathrm{L}_{+}^{k}=\mathrm{L}^{k}$, $\mathrm{L}_{k}^{\star-}=%
\mathrm{L}_{k}$, $\mathrm{L}_{k}^{\star i}=\mathrm{L}_{i}^{k\dagger}$ as 
\begin{equation*}
\mathrm{d}\hat{\varrho}\left( t\right) +\left( \mathrm{K}\hat{\varrho }%
\left( t\right) +\hat{\varrho}\left( t\right) \mathrm{K}^{\dagger }-\mathrm{L%
}^{j}\hat{\varrho}\left( t\right) \mathrm{L}_{j}\right) \mathrm{d}t=\left( 
\mathrm{S}_{k}^{j}\hat{\varrho}\left( t\right) \mathrm{S}_{j}^{\dagger i}-%
\hat{\varrho}\left( t\right) \delta_{k}^{i}\right) \mathrm{d}A_{i}^{k}
\end{equation*}
\begin{equation*}
+\left( \mathrm{S}_{k}^{j}\hat{\varrho}\left( t\right) \mathrm{L}_{j}-%
\mathrm{K}_{k}\hat{\varrho}\left( t\right) \right) \mathrm{d}%
A_{-}^{k}+\left( \mathrm{L}^{j}\hat{\varrho}\left( t\right) \mathrm{S}%
_{j}^{\dagger i}-\hat{\varrho}\left( t\right) \mathrm{K}^{i}\right) \mathrm{d%
}A_{i}^{+},
\end{equation*}
with $\mathrm{K}+\mathrm{K}^{\dagger}=\mathrm{L}_{j}\mathrm{L}^{j}$, $%
\mathrm{L}^{j}=\mathrm{L}_{j}^{\dagger}$, $\mathrm{L}_{k}^{\dagger i}=%
\mathrm{L}_{i}^{k\dagger}$ for any number of $j$'s, and arbitrary $\mathrm{K}%
^{j}=\mathrm{K}_{j}^{\dagger}$, $\mathrm{L}_{k}^{i}$, $i,j,k=1,\ldots,d$.
This is the quantum stochastic generalization of the general form \cite%
{Lin76} for the non-stochastic (Lindblad) Master equation corresponding to
the case $d=0$. In the case $d>0$ with pseudo-unitary block-matrix $\mathrm{S%
}\mathbf{=}\left[ \mathrm{S}_{\kappa}^{\iota}\right] _{\nu=-,\circ,+}^{%
\iota=-,\circ,+}$ in the sense $\mathbf{S}^{\star }=\mathbf{S}^{-1}$, it
gives the general form of quantum stochastic Langevin equation corresponding
to the HP unitary evolution for $\psi\left( t\right) $ \cite{HuPa84}.

The nonlinear form of this decoherence equation for the exactly normalized
density operator $\hat{\rho}\left( t\right) =\hat{\varrho}\left( t\right) /%
\mathrm{Tr}_{\mathfrak{h}}\hat{\varrho}\left( t\right) $ was obtained for
different commutative It\^{o} algebras in \cite{Be90c, BaBe, Be92a}.

\end{document}